\documentclass[fleqn,usenatbib]{mnras}

\usepackage{newtxtext,newtxmath}

\usepackage[T1]{fontenc}

\DeclareRobustCommand{\VAN}[3]{#2}
\let\VANthebibliography\thebibliography
\def\thebibliography{\DeclareRobustCommand{\VAN}[3]{##3}\VANthebibliography}

\usepackage{graphicx}	
\usepackage{amsmath}	
\usepackage{amssymb}	

\title[DM velocity anisotropy to the cluster edge]{Estimating the dark matter velocity anisotropy to the cluster edge}

\author[J. Svensmark et al.]{
Jacob Svensmark,$^{1}$\thanks{E-mail: jacob.svensmark@nbi.ku.dk}
Steen H. Hansen,$^{1}$
Davide Martizzi,$^{1}$
Ben Moore,$^{2}$
Romaine Tessier$^{2}$
\\
$^{1}$DARK, Niels Bohr Institute, University of Copenhagen, Lyngbyvej 2, 4. sal, 2100 Copenhagen \O, Denmark\\
$^{2}$Institute for Computational Science, University of Zurich, CH-8057 Zurich, Switzerland
}

\date{Accepted XXX. Received YYY; in original form ZZZ}

\pubyear{2019}

\begin{document}
\label{firstpage}
\pagerange{\pageref{firstpage}--\pageref{lastpage}}
\maketitle

\begin{abstract}
Dark matter dominates the properties of large cosmological structures such as galaxy clusters, and the mass profiles of the dark matter have been measured for these equilibrated structures for years using X-rays, lensing or galaxy velocities. A new method has been proposed, which should allow us to estimate a dynamical property of the dark matter, namely the velocity anisotropy. For the gas a similar velocity anisotropy is zero due to frequent collisions, however, the collisionless nature of dark matter allows it to be non-trivial. Numerical simulations have for years found non-zero and radially varying dark matter velocity anisotropies. Here we employ the method proposed by Hansen \& Pifaretti (2007), and developed by H\o st et al. (2009) to estimate the dark matter velocity anisotropy in the bright galaxy cluster Perseus, to near 5 times the radii previously obtained. We find the dark matter velocity anisotropy to be consistent with the results of numerical simulations, however, still with large error-bars. At half the virial radius we find the velocity anisotropy to be non-zero at 1.7$\,\sigma$, lending support to the collisionless nature of dark matter.
\end{abstract}




\section{Introduction}

The global dynamics of the expanding universe is dominated by two
invisible components, namely dark matter (DM) and dark
energy~\citep{planck18}. In addition there exist
independent gravitational observations of dark matter on smaller
scales~\citep{clowe2006}.  Despite the importance of dark
matter in structure formation, we still have only limited knowledge
about its fundamental properties.

From a basic point of view, DM is constituted of fundamental
particles, characterized by their mass and interactions with other
particles.  These parameters can be tested through astronomical
observations as well as in terrestrial experiments.  Typically,
cosmological observations measure a combination of these. For
instance, if DM particles interact with photons, structure formation
will be affected through the ratio of the interaction cross section
and the DM particle mass, $\sigma_{\gamma - \mathrm{DM}}/M_{\mathrm{DM}}$
\citep{bloehm2002,hinshaw2013}. Similar constraints
can be obtained for DM self-scattering or for various annihilation
channels (for a list of references, see
e.g. \cite{zavala2013,liu2016}). A range of
accelerator and underground detector ``null'' observations have
provided limits on the DM mass and interaction rates, e.g., from CMS,
ATLAS, DarkSide-50, LUX \citep{lowette2014,
aad2015,agnes2015,faham2014}.
Basically these constraints indicate that the DM has only very limited
interactions besides gravity.

Structure formation has been thoroughly investigated for many years
using numerical simulations in a cosmological setting. The resulting
structures include galaxies and clusters at various stages of
equilibration.  These simulations have revealed that the DM density
profile, $\rho(r)$, of individual cosmological structures changes from
having a fairly shallow profile in the central regions, $\gamma_\rho =
d\, {\rm log}{\rho}/ d {\rm log} r \approx -1$, to a much steeper fall
off in the outer regions, $\gamma_\rho \approx -3$
\citep{navarro2010}.

For the largest equilibrated structures like galaxy clusters there is
fair agreement between the numerical predictions and observations
concerning the central steepness \citep{pointecouteau2005,vikhlinin2006}.  
However, for smaller structures like galaxies
or dwarf galaxies, the observations have indicated that the central region
has a shallower density profile than seen in numerical simulations
\citep{salucci2007,gilmore2007}, and it is not entirely resolved whether 
this difference is because of significant self-interaction of the
DM, or because of stellar, black hole, or supernova effects. 
The majority of recent state of the art
simulations employing cold dark matter models with baryonic effects and
stellar feedback tend to agree with observations 
\citep{amorisco2014,santos-santos2017,dutton2018,wheeler2018,benitez-llambay2018,
wetzel2016,bullock2015,teyssier2013}, however some still do not find
cores using this approach \citep{bose2018}. Some efforts have been made
with alternative dark matter models, but the results are thus far not 
fully conclusive \citep{dicintio2017,fitts2018,gonzales-sameniego2017}.
 
Observationally it is very difficult to determine other properties of
the DM structures besides the density profile. The density profile is
a static quantity (not involving velo\-ci\-ty), which arises from the
zeroth moment of the Boltzmann equation (i.e. mass conservation). The
first moment of the Boltzmann equation instead relates to momentum
conservation, and here appears the first dynamical properties in the 
shape of the so-called dark matter velocity anisotropy. 

The principal purpose of measuring the dark matter velocity anisotropy
is to improve our knowledge of dark matter.  The value of the velocity 
anisotropy depends on the magnitude of the dark matter self-interactions 
\citep{brinckmann2018}, and hence a precise measurement of the velocity 
anisotropy in the inner halo region should allow us to constrain the 
dark matter collisionality.  Furthermore, it has been suggested from 
theoretical considerations that there should be a correlation between 
the dark matter velocity anisotropy and the total mass
profile \citep{hansen2009}, and a future accurate
measurement of both would allow testing this prediction.  Finally, it
is possible that the velocity anisotropy in the outer cluster regions
could depend on cosmological parameters, even though, to our
knowledge, this has not yet been thoroughly investigated. Hopefully an
investigation like the one we are presenting here, will inspire
simulators to make such an investigation.

We will in this paper attempt to measure the dark matter velocity 
anisotropy.  The technique we
use is based only on the observation of hot X-ray emitting gas, and it
uses the combined analysis of both the gas equation (hydrostatic
equilibrium) and the DM equation (the Jeans equation). 
Measurements and analyses of the X-ray emitting hot gas have improved significantly
over the last decades, and we will use the recent observation of
Perseus, which extends up to the virial
radius \citep{simionescu2011, urban2014}. Here we define the virial
radius as the radius where the enclosed density is 200 times the
critical density of the universe, $r_{\rm vir} = r_{200}$. 
This approach of measuring velocity anisotropy contains 
the radial velocity dispersion of the DM as a degeneracy, and thus the resulting 
velocity anisotropy should be viewed as a check of the consistency of data with the DM model of the
simulation that it inherently relies on.

Probing this dynamical dark matter property was
originally suggested in \cite{hansen2007}, however, the first
reliable estimate was made by stacking 16 galaxy clusters
\citep{host2009}. This stacked cluster measurement extended
to approximately $0.85$ times $r_{2500}$. Thus, in this paper we
will extend this estimate by approximately a factor 5 in radius. 
The following sections outline how this is done through our implementation 
of this method in the context of the Perseus observations.

\section{Hydrostatic gas and equilibrated DM}\label{sec:hydrostatic_eq}

The conservation of momentum for a fluid leads to the Euler equations,
which for spherical and equilibrated systems reduce to the equation
of hydrostatic equilibrium
\begin{equation}
\frac{G M(r)}{r} = -\frac{k_b T_{\mathrm{gas}}}{m_p \mu_{\mathrm{gas}}} 
\left( \frac{\partial{\rm ln}\rho_{\mathrm{gas}}}{\partial{\rm ln}r} +
\frac{\partial{\rm ln}T_{\mathrm{gas}}}{\partial{\rm ln}r} \right) \, .
\label{eq:he}
\end{equation}
This equation simply states that when we can measure the gas
temperature and gas density (all quantities on the r.h.s. of this
equation) then we can derive the total mass profile.  From the total
mass profile one can then derive the dark matter density profile.  The
gas properties are typically observed through the X-ray emission from
bremsstrahlung, and this X-ray determination of the dark matter
density profile is very well established \citep{sarazin1986}. 
Alternatively, both density and temperature profiles can in principle
be measured separately through the Sunyaev-Zeldovich effect.

Let us now consider the dynamical equation for the dark matter.  The
dark matter is normally assumed to be collisionless, and hence the
fluid equations do not apply. Instead one starts from the
collisionless Boltzmann equation. The first moment of the
collisionless Boltzmann equation leads to the first Jeans equation,
which for spherical and fully equilibrated systems reads \citep{bt87}
\begin{equation}
\frac{G M(r)}{r} = - \sigma_r^2 \left( 
\frac{\partial{\rm ln}\rho}{\partial{\rm ln}r} + 
\frac{\partial{\rm ln}\sigma_r^2}{\partial{\rm ln}r} + 2 \beta \right)\,.
\label{eq:jeans}
\end{equation}
If we look at the r.h.s. of the Jeans equation, we see that there are
3 quantities: the dark matter density, $\rho(r)$, the radial velocity 
dispersion, and the velocity anisotropy
\begin{equation}
\beta \equiv 1 - \frac{{\sigma_\theta}^2 +{\sigma_\phi}^2 }{2{\sigma_r}^2} \, ,
\label{eq:beta}
\end{equation}
where $\sigma_r^2$, $\sigma_\theta^2$, and $\sigma_\phi^2$ are the
velocity dispersions of dark matter along the radial, polar, and
azimuthal directions respectively.

We can measure the total mass and the DM density from the equation 
of hydrostatic equilibrium.  That means that if we wish to 
determine the velocity anisotropy, then we must find a way to 
measure the radial velocity dispersion of the dark matter, 
$\sigma_r^2$. To that end we will need assistance from numerical 
simulation, which we will explain in detail below. The conclusion 
will be that we can map the gas temperature to the DM velocity 
dispersion. Thus the estimation of the dark matter velocity 
anisotropy depends on the ability of numerical simulations to 
reliably map between gas temperature and DM dispersion.

The dark matter particles are normally assumed to be collisionless,
and hence the halos of dark matter will never achieve a thermal
equilibrium with Maxwellian velocity
distributions. Therefore the DM cannot formally be claimed to have a
``temperature''. However, for normal collisional particles there is a
simple connection between the thermal energy of the gas and the
temperature, and we use a similar terminology for dark matter, and
hence discuss its ``temperature'' as a measure of its local kinetic
energy.
\begin{equation}
T_{\mathrm{DM}} \equiv \frac{m_p \mu_{\mathrm{DM}}}{3 k_b} \sigma_{\rm DM}^2 \, ,
\label{darktemp}
\end{equation}
where the total dispersion is the sum of the three one-dimensional 
dispersions
\begin{equation}
\sigma_{\mathrm{DM}}^2 \equiv \sigma_r^2 + \sigma_\theta^2 +  \sigma_\phi^2 \, .
\label{eq:sigtot}
\end{equation}

Since the dark matter and gas particles inside an equilibrated
cosmological structure experience the same gravitational potential,
then we should expect the gas and DM temperatures to be approximately
equal~\citep{hansen2007}.  Later analyses have
shown~\citep{host2009, hansen2011} that the ratio
of DM to gas temperatures
\begin{equation}
\kappa \equiv \frac{T_{\mathrm{DM}}/\mu_{\mathrm{DM}}}{T_{\mathrm{gas}}/\mu_{\mathrm{gas}}} \, ,
\label{eq:kappa}
\end{equation}
is a slowly varying function of radius, always of the order unity.

\section{DM velocity anisotropy from observables}

The Jeans equation can be rewritten as
\begin{equation}
\beta = - \frac{1}{2} \, \left(  \frac{\partial{\rm ln}\rho}{\partial{\rm ln}r} 
+ \frac{\partial{\rm ln}\sigma_r^2}{\partial{\rm ln}r}  + \frac{G M(r)}{r \sigma_r^2} \right) \, .
\label{eq:betagje}
\end{equation}

As discussed above, by measuring the gas temperature and density, 
the equation of hydrostatic equilibrium, Eq.~(\ref{eq:he}), 
gives us the total mass profile. In addition this allows us to derive the
DM density profile, $\rho = \rho_{\rm tot} - \rho_{\rm gas}$. 
Thus we only need an expression for the radial velocity dispersion 
of the dark matter, $\sigma^2_{r}$.

Combining the definitions in Eqs.~(\ref{eq:beta}, \ref{eq:sigtot}, \ref{eq:kappa}) gives
\begin{equation}\label{eq:beta_calc}
2 \sigma_r^2 \beta = 3 \sigma_r^2 - 3 \frac{k_b T_{\mathrm{DM}}}{m_p \mu_{\mathrm{DM}}} \, ,
\end{equation}
which allows us to rewrite the Jeans equation as
\begin{equation}\label{eq:sig_r}
\sigma _r ^2\left(
\frac{d\,\ln \rho_{\mathrm{DM}}} {d\,\ln r}+
\frac{d\,\ln \sigma _r ^2      } {d\,\ln r}
+3 \right) = \psi(r).
\end{equation}
where the quantity 
\begin{equation}
    \psi(r)=\left(\frac{3 k_b}{m_p} \frac{T_{\mathrm{DM}}}{\mu_{\mathrm{DM}}} - \frac{G M(r)}{r}\right)
\end{equation}
contains quantities from the X-ray observables using also $\kappa$ from equation \ref{eq:kappa}. The solution to this differential equation depends on the boundary
condition on $\sigma_r$. Here we assume that $\sigma^2_r(0)=0$. 

In this way we have all the quantities on the r.h.s. of
Eq.(\ref{eq:betagje}), and thus a measure of $\beta$.

\section{Numerical simulation and parametrizing \texorpdfstring{$\kappa$}{kappa}}
\begin{figure*}[ht!]
  \centering
  \includegraphics[width=0.46\textwidth]{{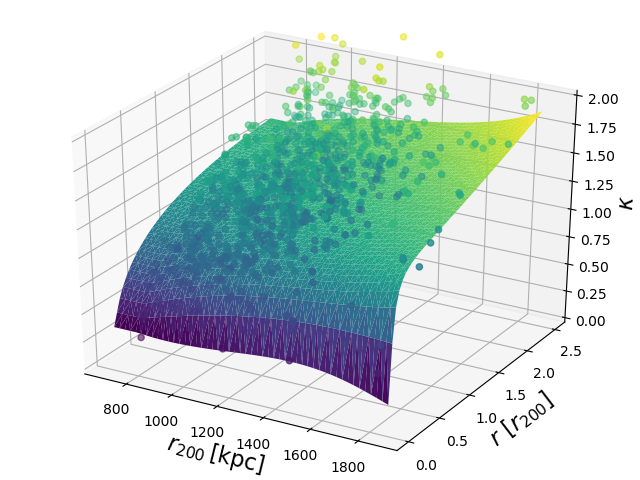}}
  \includegraphics[width=0.46\textwidth]{{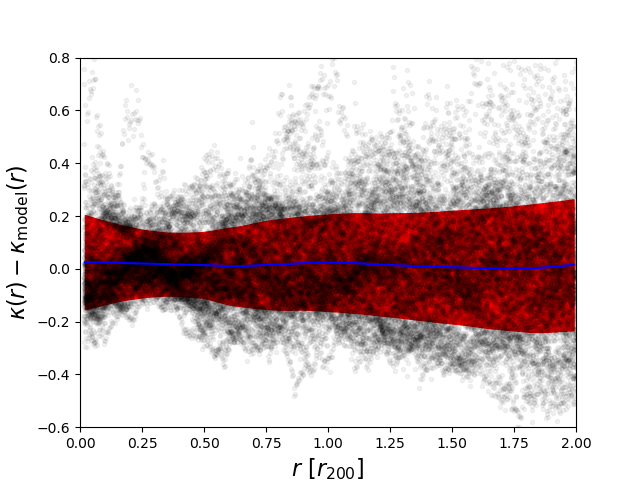}}
  \caption{$\kappa$-profiles for the 51 clusters of the RAMSES simulation, as 
  function of radius and $r_{200}$. Left panel shows a 2D smoothing spline to the 
  profiles, and right panel shows the 1$\sigma$ 
  scatter contours of the residuals collapsed along the $r_{200}$ direction. The 
  smoothed surface, and scatter profile serves as parametrization for $\kappa(r,r_{200})$ 
  profiles for observations.}
  \label{fig:kappa}
\end{figure*}

The energy-argument that the DM dispersion should be approximately equal to the gas temperature ($\kappa\approx1$) in relaxed gravitating structures has a long history \citep{sarazin1986}. The anticipation that $\kappa$ may change significantly when gas is cooling was investigated by \cite{hansen2011}, where $\kappa$ was extracted for Milky Way like galaxies as a function of redshift (where cooling is extremely much more significant than in cluster outskirts). There it was found that as long as the high-density/low-temperature component of the gas is removed, then $\kappa$ remains close to unity around $z=0$.

Here we take possibly the most modern approach to gas cooling and other radiative processes in simulation to extract $\kappa$. We chose to use a simulation with the AMR code RAMSES 
\citep{teyssier2002}, which uses flat $\Lambda$CDM
cosmology with cosmological constant density parameter $\Omega_\Lambda
= 0.728$, matter density parameter $\Omega_m = 0.272$ of which the
baryonic density parameter is $\Omega_b = 0.045$, power spectrum
normalization $\sigma_8 = 0.809$, primordial power spectrum index $n_s
= 0.963$, and current epoch Hubble parameter $H_0 = 70.4\,\mathrm{km}/\mathrm{s}/\mathrm{Mpc}$. 
To identify large galaxy clusters, 
the simulation was initially run as a dark-matter-only
simulation with comoving box size $144\,\mathrm{Mpc}/h$ and particle mass $m_{\mathrm{DM}}
= 1.55 \cdot 10^9\,M_{\odot}/h$. Here $h$ is the dimensionless Hubble
parameter, defined as $h = \frac{H_0}{100\,\mathrm{km}/\mathrm{s}/\mathrm{Mpc}}$. 
After running the dark
matter only simulation, 51 cluster sized haloes with total
masses above $10^{14}\,M_{\odot}/h$ were identified and 
resimulated including the baryonic component, with dark matter particle
mass $m_{\mathrm{DM}} = 1.62 \cdot 10^8\,M_{\odot}/h$ and baryonic component
mass resolution of $3.22 \cdot 10^7\,M_{\odot}$. The 51 resimulation
runs implemented models of radiation, gas cooling, star formation,
metal enrichment, supernova and AGN feedback, and were
evolved to $z=0$. A detailed description of the
simulation can be found in \cite{martizzi2014}. 

For the 51 clusters, the $\kappa$ profile can be calculated in spherical bins 
according to equation \ref{eq:kappa}. Since
all quantities contained in $\kappa$ depend on the cluster size, we calculate a
2D smoothing spline surface for the 51 $\kappa(r,r_{200})$ profiles, such that 
given $r_{200}$ for a cluster, $\kappa(r)$ can be retrieved (left panel of figure 
\ref{fig:kappa}). The error associated with using this $\kappa$ function is 
approximated from the residual after collapsing it in the $r_{200}$ direction 
(figure \ref{fig:kappa} right panel), and we find no strong correlation or systematics within these residuals. The resulting 1$\,\sigma$ standard deviation 
profile can then be taken into account when estimating $\beta(r)$. When we compare with older numerical techniques, such as the use of \texttt{GADGET-2} \citep{kay2007} (previously used to extract $\kappa$ \citep{host2009}), then we find that the resulting $\kappa$ profiles are in fair agreement with each other, within the error-bars.

Herein lies the core of the method, but notably also the reason 
why resulting $\beta$ profiles cannot be called de facto measurements. Rather they 
are consistency checks with the DM model employed in the simulation that produces 
the $\kappa$ relation. 
The Jeans equation assumes DM to be collisionless, as (in this case) does the simulation
that produces $\kappa$, however should this assumption not be valid, a measured
$\beta$ profile might not be consistent with those of simulations. From an observational 
point of view, the $\kappa$-parametrisation makes good sense, 
as measurements of the mass profile and thus $r_{200}$ of galaxy clusters are 
independent of $\kappa$. Thus by observing the properties of the hot X-ray 
emitting gas we can obtain $\kappa(r)$ using our parametrization, and from this 
calculate $\sigma_r^2(r)$ from eq. \ref{eq:sig_r}
and finally $\beta(r)$ from eq. \ref{eq:beta_calc} and \ref{eq:kappa}, assuming 
that the gas is fully equilibrated. The next 
section is dedicated to reinforcing the soundness of this assumption in the 
observations that we choose to analyze.

\section{Excluding infeasible sectors from analysis}\label{sec:selection}

One of the core assumptions in deriving massprofiles from the X-ray signal
in galaxy clusters is that of hydrostatically equilibrated
gas. This excludes a large block of potential cluster targets for study, as merging 
and other irregularities causes a failure to meet this demand. Previous studies show
how cluster merging can cause cold front and sloshing in the hot baryonic gas, and
phenomena that show up in derived X-ray profiles as unequilibrated features \citep{markevitch2007}.
Data quality has however heightened through the last decades, and so has the frequency of attempts at 
solving this issue through data selection - considering only sections of the 
observational plane which best meets the assumptions of equilibrium. This makes sense if material 
falling onto an equilibrated structure is small enough to only disturb
equilibrium locally, or that the in-falling material has not yet had time to perturb
the larger equilibrated cluster in its entirety.

In order to develop and test a method of measuring $\beta$ to high radii
through data selection in the observational plane of galaxy clusters, we
construct a mock observational catalogue from the RAMSES re-simulated clusters. 
Radial profiles of all quantities relevant for the 
present work has been extracted. We select a sub-sample of 10 randomly selected clusters in 
range $14 < \log{m_{200}} < 15$ where half of the clusters in the sample have been classified 
as globally relaxed, following the criterion of Martizzi and Agrusa 2016 \citep{martizzi2016}.
We choose a line of sight through them at random and divided each observational 
plane of the 10 clusters into 8 equiangular sectors.
The motivation for this division originates in the Perseus X-ray observations 
analyzed in the next sections. Perseus
is precisely observed along 8 arms at evenly spaced angles, yielding 8 gas density 
and temperature profiles in total for the cluster. Previous analysis of the 8 arms of Perseus
has shown signatures of cold-fronts and sloshing of the gas in 5 of the 8 arms 
suggesting that they are suboptimal for calculations assuming hydrostatic equilibrium. 
The remaining 3 are shown to be more relaxed. In the following we take a similar 
approach selecting just the sectors that are equilibrated. We conclude that by 
deselecting the most deviant sectors we can reduce the scatter in our final 
measurements of $\beta$ within the RAMSES clusters.

For each of the sectors in the observational plane, 3D radial profiles from 
spherical shells were extracted for 
the gas and dark matter component, using only the particles that in projection are 
contained inside of that sector. Each sector can then be analyzed separately, and thus 
parts of the observed cluster that displays unequilibrated features may be excluded 
from our calculation of $\beta(r)$. With this in hand, we estimate the statistical 
reward in removing sectors from analysis, and use this in an attempt to reduce 
the errorbars of $\beta(r)$ for actual X-ray data.

\begin{figure*}
    \hspace*{-1.3em}
  \centering
  \includegraphics[width=0.96\textwidth]{{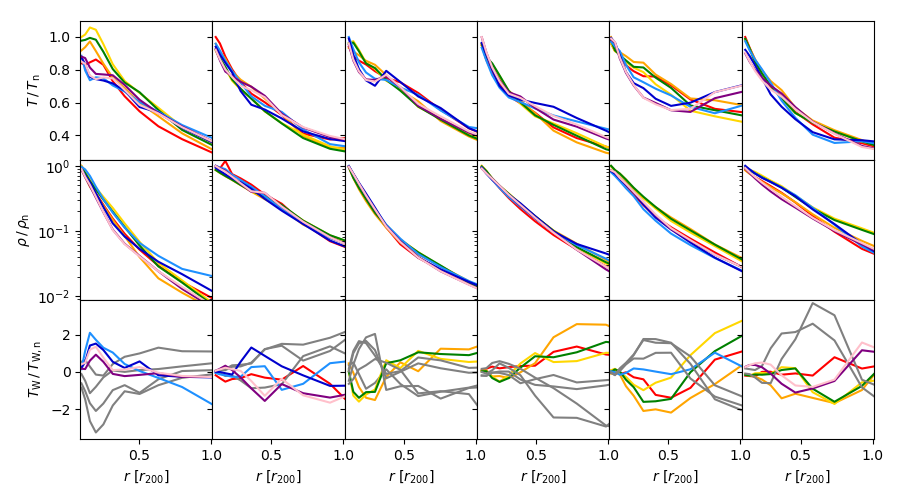}}
  \caption{Temperature profiles (top row), gas density profiles (middle row) 
  and weighted temperature variation profiles (bottom row) for the six resimulated 
  RAMSES clusters considered for analysis. Each color represents one of the 8
  sectors within the cluster. Neighbouring sectors are colored in the following
  order: Red, orange, yellow, green, light blue, dark blue, purple, pink. The 
  gray profiles in the bottom row indicates the sectors that were excluded from
  analysis due to their profile as discussed in the main text. Note that $T$, $\rho$ 
  and $T_{W}$ profiles are scaled by a constant $T_\mathrm{n}$, $\rho_\mathrm{n}$ and
  $T_\mathrm{W,n}$, which differs within each cluster, in order to compare profile 
  shapes between clusters in this figure.}
  \label{fig:massweight}
\end{figure*}

Figure \ref{fig:massweight} displays gas temperature and density profiles for 6 
of the 10 resimulated galaxy clusters for which we have 3D data. The remaining
4 clusters displayed enough unequilibrated features in the gas component
that they were unfit to consider for further analysis, due to strongly inconsistent profiles
across the 8 sectors of a cluster, or a steep incline in the central temperature profiles. 
It should be noted, that only three of the six
clusters remaining were categorized as virialized using the criteria of \cite{martizzi2016}.  
Each column of figure \ref{fig:massweight}
represent one of the six clusters, and the colored profiles represent individual
sectors. Top panels show the gas temperature scaled by a constant, and the central 
panels show the gas density also scaled. From these two quantities alone it is clear 
that sectors within a single cluster displays some variance, which is natural to 
expect from  a 3D numerical simulation, but also a reflection of the hydrostatic 
equilibrium assumption not being perfectly true. For the remainder of this section
we will discuss which directions appear less equilibrated than others, considering 
purely the observational gas profiles.

For some of the clusters, looking 
only at the top two rows of figure \ref{fig:massweight} it is not always
easy to tell whether parts of the cluster are effected by some disturbing element.
The 5th column has a pretty clear signature in the temperature profiles that
something is disturbing its equilibrium in the dark blue, purple and pink sectors.
For the cluster in the 6th column, deviations from 
hydrostatic equilibrium are more subtle. 
Herein lies an observational challenge, and in an attempt to  enhance the 
visibility of such subtle differences between sectors within a single galaxy 
cluster, we combine the two measurable quantities $\rho_{\mathrm{gas}}$ and
$T_{\mathrm{gas}}$ into a single measure.

The bottom panels of figure \ref{fig:massweight} show this combination in the form
of a weighted temperature variation profile, 
\begin{equation}\label{eq:massweight}
    T_{\mathrm{W}} = \rho_{\mathrm{gas}} r^2 \left( 1-\frac{T_{\mathrm{gas}}}{\overline{T_{\mathrm{gas}}}}\right),
\end{equation}
where $\overline{T_{\mathrm{gas}}}$ is the mean temperature profile of the entire cluster. This observationally available construct emphasizes in some cases distinct groups of sectors, and by comparing these to the gas and temperature profiles of the same cluster, we try to deselect those that are least consistent with the overall trend of the cluster. Here bumps i.e. cold fronts in the temperature and density profile are features we look for \citep{markevitch2007}. In the case of well behaved clusters, this of course is less obvious, and arguably data selection may also have less of an effect.
In column 3 of figure \ref{fig:massweight} the density is smooth, but the 
dark blue, light blue, pink, purple and red directions have a bump in the temperature profile. The
$T_{\mathrm{W}}$ profiles show two groups that clearly differ from each other. Based on 
the irregular bump in the $T_{\mathrm{gas}}$ profile and the separation in the $T_{\mathrm{W}}$ figure we 
include only the green, orange and yellow directions, and indicate the deselected sectors 
with the grey color in the bottom panel. In
column 6 the two blue sectors stand out slightly in the $T_{\mathrm{gas}}$-profile, but 
more profoundly in the $T_{\mathrm{W}}$ profile, and are thus removed from analysis.
Column 2 displays a very smooth cluster, and it is less clear which (if any) directions 
should be removed. We deselect the green, yellow and orange directions as they tend to lie
slightly high for the high radii of primary interest for this paper. The cluster in the first 
column shows a kink towards its inner parts. 
Here, the red, orange, green and yellow profiles deviate largely within 0.7$r_{200}$ from the
remaining four sectors, and are thus removed from analysis. The $T_{\mathrm{W}}$ profiles of column 4
shows three main groups, substantiating what is otherwise hard to see in the $T_{\mathrm{gas}}$ 
profiles alone. We remove from analysis the dark blue, light blue, purple and pink and keep the 
most coherent 4 sectors as indicated in the figure.

In some clusters, such as the one in column 5 of figure
\ref{fig:massweight}, signatures of a ''cold-front'' is visible in the
sectors represented by the dark-blue, pink and purple profiles, which are 
consequently removed from analysis. These may be
caught by performing an analysis of the X-ray data analogous with the one in 
\cite{urban2014}, and in this case both approaches would possibly single out the 
same sectors. The present exclusion process however singles out 
some features that are not predominantly cold-front related, and as such provides 
a different approach to determining which parts of a galaxy cluster that are not in
equilibrium. 

From the weighted temperature variation profiles in combination with the raw density and 
temperature profile, we have identified up to five sectors within each cluster that 
deviate substantially from the more relaxed conditions, and are now ready to 
calculate the velocity anisotropy parameter $\beta$.

\section{Non-parametric fitting and MC resampling to data}\label{sec:practice}
\begin{figure}
    \hspace*{-1.3em}
  \centering
  \includegraphics[width=0.50\textwidth]{{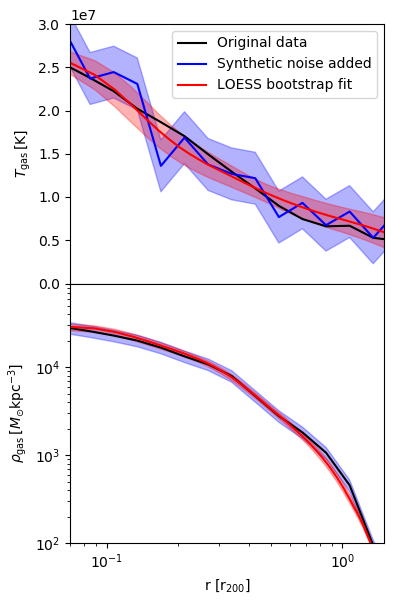}}
  \caption{Example of the temperature (upper panel) and density (lower panel) 
  LOESS fits employed in the hydrostatic equilibrium calculation for the gas 
  component, as applied to a single sector from one of the simulated clusters. Top 
  panel shows $T_{\mathrm{gas}}$ and bottom panel $\rho_{\mathrm{gas}}$.
  Black lines show the original profiles. The blue curves and 
  patches shows the profile with synthetic noise and errorbars added, and the red 
  curves shows the median and 68\% percentile of 100 bootstrap resampled 
  Monte Carlo LOESS fits to the noisy profiles.}
  \label{fig:TrhoLOESS}
\end{figure}

In order to arrive at an estimate of $\beta$, local fluctuations and 
measurement uncertainties are necessary to take into account. We employ 
non-parametric locally estimated scatter plot smoothing (LOESS) fitting to 
the gas density and temperature measurements in order to smooth out local 
variations \citep{msir2011}. In this way we manage to avoid imposing an analytical profile 
to our data. This yields a fitted curve and a 1$\sigma$ standard 
deviation profile in addition. Any fit, including this one, is of course subject to
a level of arbitrarity in the choice of function, and for non-parametric fits
some choice of smoothing parameter and algorithm. In this case, we let the LOESS smoothing
parameter be determined by a generalized cross-validation technique \citep{fANCOVA2010}, 
which adapts to the data in question. Doing this, we obtain a smooth profile, that neglects 
local bumps and wiggles, however allows for larger scale variation, rather than forcing 
it to follow a strict parametric form. An example of a cross-validated LOESS 
fit to the gas temperature profile is seen is seen in figure 
\ref{fig:TrhoLOESS} for a single sector in one of the galaxy clusters within the 
RAMSES simulation. In this example the temperature profile from the simulated data 
is without uncertainty and smooth in 
comparison with realistic measurements of the 3D temperature profiles of galaxy 
clusters. Therefore Gaussian errors of $3\times 10^6\,$K and error bars of the 
same magnitude are added. The red band of figure \ref{fig:TrhoLOESS} represents the 68\% 
percentile of 100 LOESS fit to the noisy data using a resampling 
technique assuming a Gaussian distribution, and recreates the original 
temperature profile in black reasonably well. The same goes for the 
$\rho_{\mathrm{gas}}$ in the bottom panel, though this is to be expected as uncertainties 
in typical gas density profiles are small compared to the temperature measurements. 
As explained in section \ref{sec:error} we employ a Monte Carlo resampling approach 
to obtain error estimates on the measurement of the galaxy cluster $\beta$. 
For an input smoothed $T_{\mathrm{gas}}$ and $\rho_{\mathrm{gas}}$ we proceed to 
calculate $M(r)$, $\sigma_r^2(r)$ and $\beta(r)$ through hydrostatic equilibrium 
assumptions, as shown in section \ref{sec:hydrostatic_eq}. In this process,
we fit yet another LOESS curve to both the $M(r)$, $\rho$ and $\sigma_r^2(r)$ profiles,
to neglect the smaller bumps and ripples. This comprises the drawback of not 
assuming and fitting e.g. well behaved power law functions to the raw hydrostatic
data. However, the multiple non-parametric fits do allow a degree 
(as controlled by the cross-validation mentioned above) of ripple that would 
otherwise not be seen in the parametric form, and in this respect the approach is
arguably preferable. 
In the next section we can begin the process of Monte Carlo resampling
$\rho_{\mathrm{gas}}$, $T_{\mathrm{gas}}$ and $\kappa$ to arrive at a final estimate 
of $\beta$ and its uncertainty from a number of sectors within a single
galaxy cluster.

\section{Errors in estimating  \texorpdfstring{$\beta$}{beta}}\label{sec:error}
\begin{figure}
  \centering
  \includegraphics[width=0.48\textwidth]{{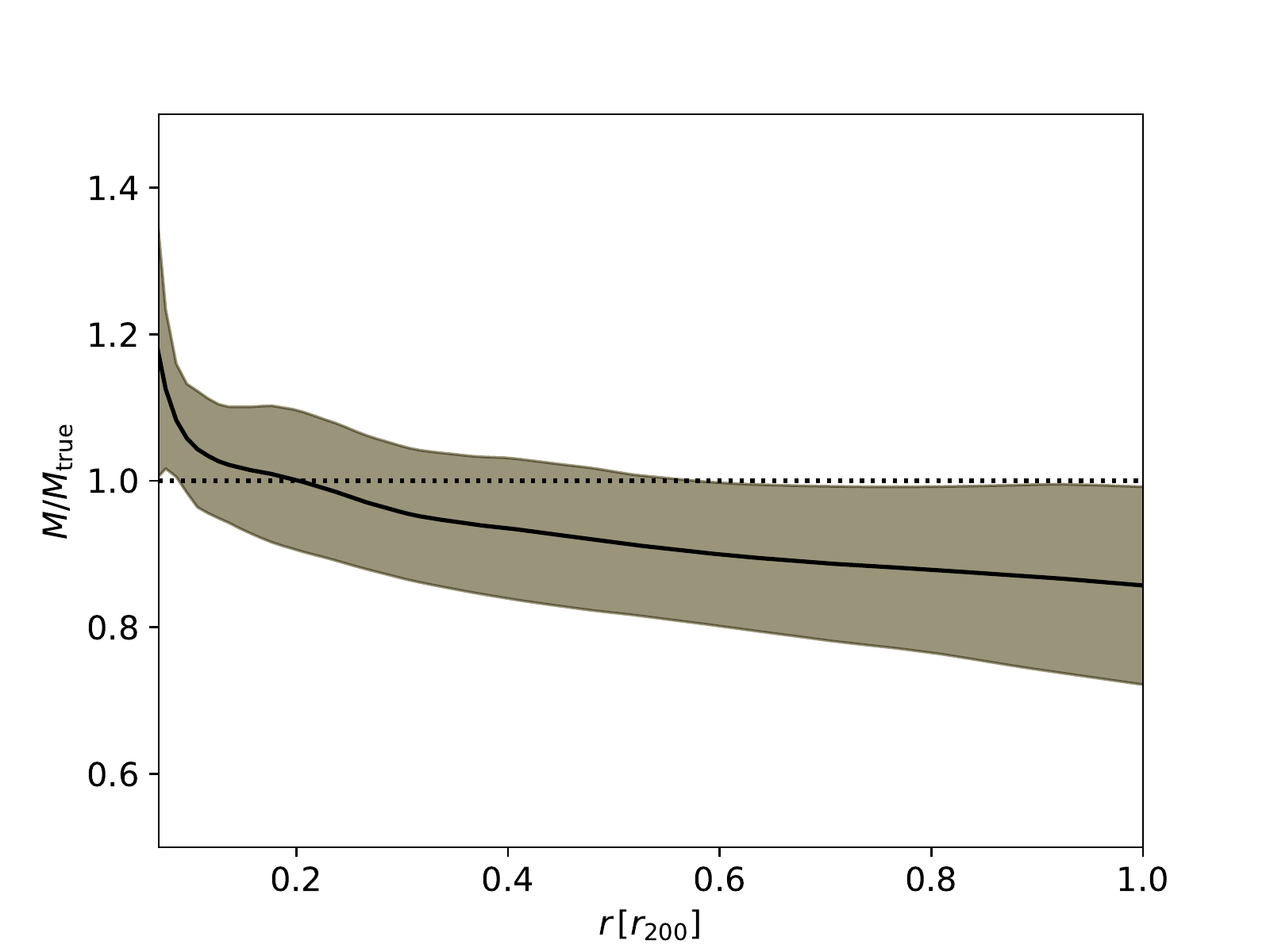}}
  \caption{The mass profiles from hydrostatic equilibrium for all sectors of the 6 
  RAMSES clusters relative to the true mass profile of the cluster they belong to. The black
  curve represents a LOESS fit to the individual profiles, and the grey band 
  represents the 1$\,\sigma$ standard deviation of the profiles, as obtained from
  the generalized cross-validation technique of the LOESS fit outlined in the main text.}
  \label{fig:mass_RAMSES}
\end{figure}

\begin{figure*}
    \hspace*{-1.3em}
  \centering 
  \includegraphics[width=0.49\textwidth]{{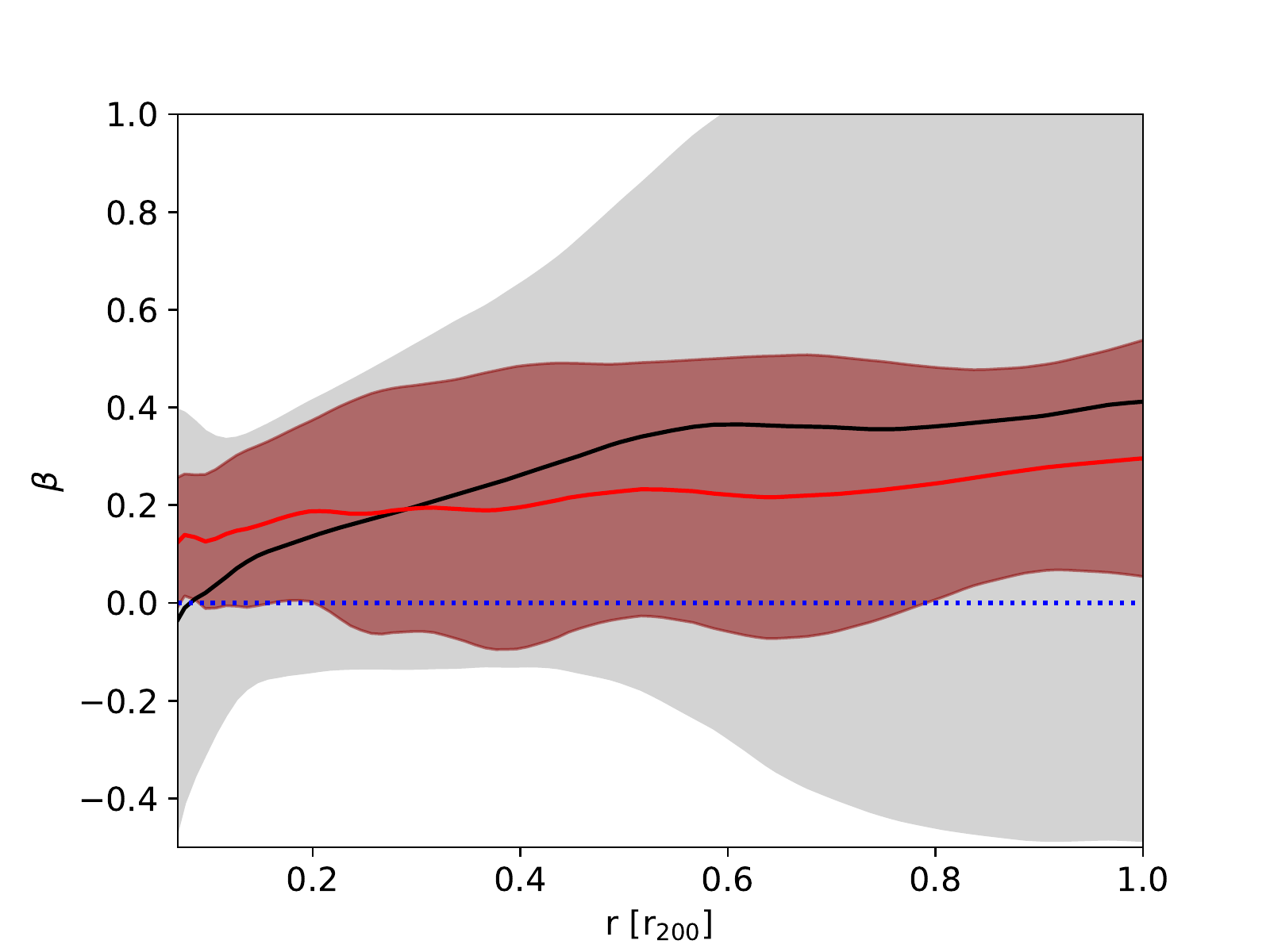}}
  \includegraphics[width=0.49\textwidth]{{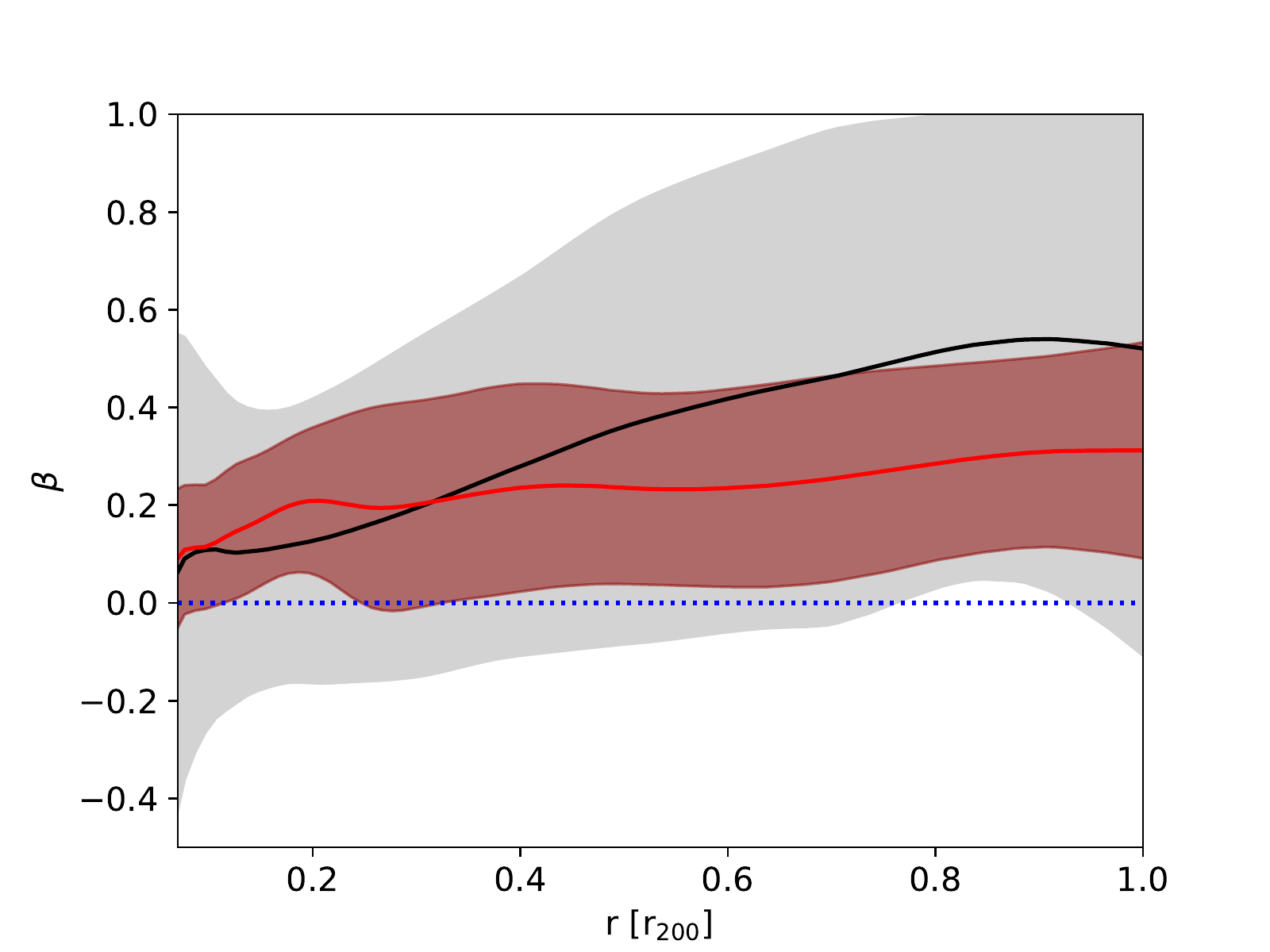}}
  \caption{$\beta$ profile for all sectors (left) and selected sectors (right) 
  for the six relaxed RAMSES clusters using interpolated 
  2D smoothing spline $\kappa$-profiles for each cluster. Red curve shows a LOESS fit to the 
  true $\beta$-profiles, and black curves to measured ones. The red narrow patches indicate 
  the 1$\sigma$ standard deviation of the true $\beta$-profiles, and the light grey wider 
  patches indicated the same but for the measured $\beta$-profiles, as obtained through the
  generalized cross validation technique of the LOESS fit outlined in the main text. Thus the
  standard deviations shown here is for a single sector of a single cluster.}
  \label{fig:beta}
\end{figure*}

Each sector of each cluster is handled individually in our analysis. The final $\beta$ of a given sector is obtained using a Monte Carlo resampling approach, which allows us to propagate measurement errors from the input X-ray profiles. For a single sector we produce a number $N_{\mathrm{MC}}$ of resamples of $T_{\mathrm{gas}}(r)$ using its measurement uncertainties. We resample complementary $\kappa(r)$ profiles and proceed to calculate $M(r)$, $\sigma_r^2(r)$ and $\beta(r)$ through hydrostatic equilibrium
assumptions, as shown in section \ref{sec:hydrostatic_eq} and \ref{sec:practice}. The final $\beta$ profile for a given sector is then the median profile of all $\beta$ from the resamples of that sector.
Our intent is to use the procedure on multiple sectors of a single cluster (or potentially even multiple clusters, though this is left for future work), and end up with a final estimate of the universal velocity anisotropy profile. We must therefore understand to what degree the procedure is biased, and how much scatter it introduces in addition to the natural scatter within cluster $\beta$ profiles. To do this, we measure the $\beta$ profile of two groups of sectors from the 6 RAMSES clusters selected in section \ref{sec:selection}: One group consisting of all 48 sectors in the 6 RAMSES clusters, and another group using only the 27 equilibrated sectors. Starting with the full set, as an intermediary step the hydrostatic massprofiles of each sector is calculated. These can be seen in figure \ref{fig:mass_RAMSES}, relative to the true mass profiles. The hydrostatic masses are just within the 1$\sigma$ 
standard deviation profile at the radii of interest, however the mean value 
under-performs between between 0 and 10\% low for growing radii similar to previous findings using other mass reconstruction techniques  \citep{gifford2013,armitage2019}. One could imagine correcting inferred masses accordingly upon measurements, however it is non-trivial how that translates into a $\beta$ correction given the simulated data we have available. For this reason we allow the hydrostatic mass measurements to under perform at these radii. Proceeding towards $\beta$, figure \ref{fig:beta} on the left shows in the red curve and red band a LOESS fit and $\sigma$ scatter of the true $\beta$ profiles for the full set of sectors. The black curve and grey band shows the same, but for the $\beta$ profiles as measured by only the gas observables of each sector. The mock measurement of $\beta$ in a sector is seen to be unbiased, with a scatter determined largely by the true scatter in $\beta$ until around 0.5$\,r_{200}$, at which point the scatter is dominated by assumptions of equilibrium breaking down. In order to lessen the scatter, only the sectors selected in section \ref{sec:selection} i.e. the second 
set of sectors was used, and their true and measured $\beta$ summarized in the right hand side of figure \ref{fig:beta}. Notably, the scatter is lower in this case because assumptions of equilibrium are better met in these selected simulated sectors. 
The grey patches in the right panel of figure \ref{fig:beta} comprises scatter in the true $\beta$ profiles of the simulated clusters, as well as additional scatter introduced by our analysis. As we proceed to calculate $\beta$ for X-ray observations of the sectors in a single real galaxy cluster, we must incorporate this scatter in the uncertainty of our measurement. How this is done depends on the amount of correlation between sectors of a single galaxy cluster. If all sectors within a single cluster are completely independent measurements of $\beta$, then the uncertainty of the joint $\beta$ profile decreases by a factor $1/\sqrt{N}$ where $N$ is the number of sectors under analysis. If on the other hand sectors within a single cluster are completely correlated, the part of the scatter that originates from natural variation in $\beta$ (red patch of figure \ref{fig:beta}) is constant with number of sectors, whereas the residual scatter (difference between grey and red scatter) in figure \ref{fig:beta}, i.e. the additional scatter introduced by the analysis framework is reduced by $1/\sqrt{N}$, assuming that the two sources of scatter are directly separable. As one extreme, we could assume that all of the sectors of a single galaxy were uncorrelated in their measurement of $\beta$, and as another we could assume complete correlation.
Now we have an estimate of the uncertainties involved in measuring $\beta(r)$ through X-ray data and assumptions of hydrostatic equilibrium, and a method for 
eliminating parts of this uncertainty by data selection. In the next section
we move to apply the technique and estimate $\beta(r)$ to the virial radius for the Perseus galaxy cluster.

\section{Perseus cluster observations in X-ray}\label{sec:perseus}
\begin{figure*}
  \hspace*{-1.3em}
  \centering
    \includegraphics[width=0.9\textwidth]{{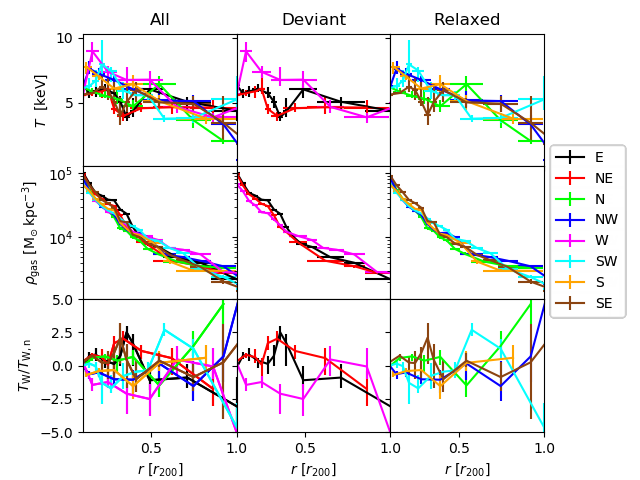}}
  \caption{Deprojected profiles for the 8 sectors of the Perseus cluster, 
  grouped in categories through columns "All", "Deviant" and "Relaxed". 
  Top row shows the gas temperature $T_{\mathrm{gas}}$, center row the gas 
  density $\rho_{\mathrm{gas}}$ 
  and bottom row the weighted temperature variation $T_{\mathrm{W}}$.}
  \label{fig:perseus_diagnostics}
\end{figure*}

The Perseus cluster is the brightest cluster in the X-ray sky, and was
observed in 85 pointings as a Suzaku Key Project, with a total
exposure time of 1.1 Ms.  The low particle background makes Suzaku
ideal for analysing cluster outskirts. These pointings were arranged
in eight arms along different azimuthal directions. For each direction
the data had point sources removed and was cleaned, 21 pointings were
used for a careful background modelling, and XSPEC was used to extract the 
deprojected temperature and density profiles~\citep{simionescu2011,urban2014}. Such careful 
treatment of the deprojection is necessary, as the calculation of $\beta$ requires 3D profiles 
of $T_{\mathrm{gas}}$ and $\rho_{\mathrm{gas}}$ to function.
For each of the eight arms, the $T_{\mathrm{gas}}$ and $\rho_{\mathrm{gas}}$ profiles can be 
seen with uncertainties in the top and middle panel of the LHS 
column in figure \ref{fig:perseus_diagnostics}.

Previous careful analyses allowed a categorization of the eight arms into
three ``relaxed'' arms showing no particular irregular behaviour,
where in particular the temperature profiles are generally decreasing
functions of radius. The other arms either show signs of large cold
fronts between 20 and 50 arcminutes form the center, or showed signs
of large scale sloshing motion of the gas~\citep{simionescu2011,
simionescu2012, urban2014}. 

In this work, we consider temperature variations relative to the mean profile, and weigh
them by $\rho r^2$ as described in equation \ref{eq:massweight}.
The $T_{\mathrm{W}}$ profiles are seen in the bottom panel of the first column in 
figure \ref{fig:perseus_diagnostics}. 
Since Perseus is already a comparatively virialised cluster, there
are not a couple of sectors that show extremely obvious deviant features
from the mean temperature profile. The western arm (magenta) shares more or less
no features with the rest, and is arguably not in equilibrium with the remaining
parts of the cluster. Beyond 0.7$\,r_{200}$ the spread of the profiles
becomes very large and the profiles deviate significantly from each other.
Within 0.7$\,r_{200}$ the Eastern (black), North Eastern (red) and to a lesser 
extent the South Eastern (brown) arms display an irregularity that Urban et. al 
conclude to be a cold front. Here we shall remove Eastern and North Eastern 
arms, the ones furthest from the mean temperature profile below 0.5$r_{200}$ and with
a general downwards tendency beyond $r_{200}$ in the $T_{\mathrm{W}}$ profiles 
(figure \ref{fig:perseus_diagnostics} bottom row central column). Instead we focus on the 
remaining 5 profiles closer to the mean below $r_{200}$ and with the upwards
tendency in the $T_{\mathrm{W}}$ profiles beyond $r_{200}$ (figure 
\ref{fig:perseus_diagnostics} bottom row 3rd column).
Then we are consistent with the methodology of the previous sections,
where the numerical clusters were considered, and arrive at the selections in the 3rd 
column of figure \ref{fig:perseus_diagnostics}. 
The middle column shows the profiles that are not included in the analysis.

This leaves us with three sets of data within Perseus: The full set, 
the sectors selected here and the ones found to be relaxed by previous 
analysis \citep{urban2014}. In the following section we examine
all three sets. Each arm is fed through 
our analysis separately, and in the end we stack the $\beta$ of each
set to obtain an overall estimate of $\beta$ from Perseus.

We do not consider radii outside the virial radius, 
where the infall motion leads to departure from hydrostatic 
equilibrium~\citep{falco2013,albaek2017}.

\begin{figure}
  \hspace*{-1.3em}
  \centering
    \includegraphics[width=0.47\textwidth]{{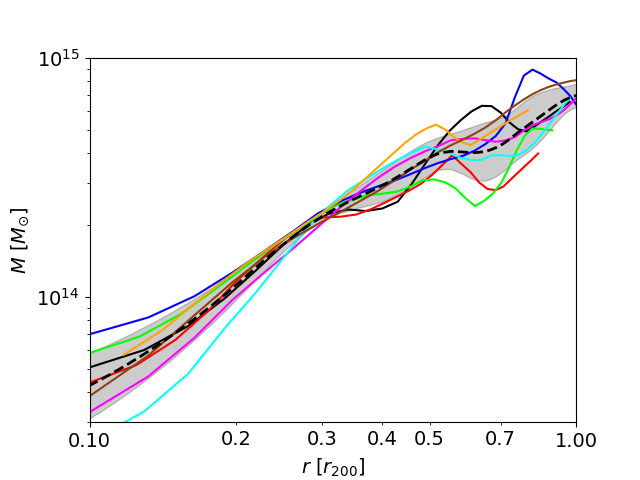}}
    \includegraphics[width=0.47\textwidth]{{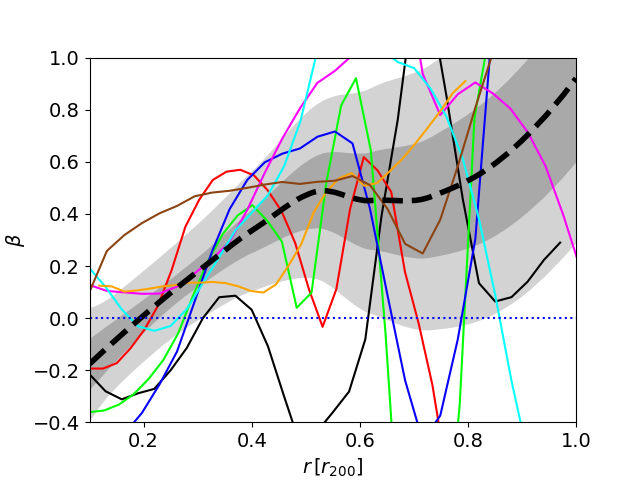}}
  \caption{Left: Mass profiles for the 8 sectors of Perseus as 
  obtained from hydrostatic equilibrium. Gray band shows the 1$\sigma$ 
  spread of the individual profiles. Note the logarithmic $r$-axis. 
  Right: Calculated $\beta(r)$ profiles 
  for the same directions arms. Dark grey band represents the uncertainty of 
  the mean $\beta$ profile based on the spread of the Perseus 
  sample ($\frac{\sigma}{\sqrt{N}}$ where $N$ is number of sectors), 
  and and light grey is the added uncertainty of the mean based 
  on the standard deviation of the RAMSES full set i.e. the grey area 
  of the LHS panel of figure \ref{fig:beta}.}
  \label{fig:perseus_mass}
\end{figure}

\begin{figure}
  \centering
    \includegraphics[width=0.47\textwidth]{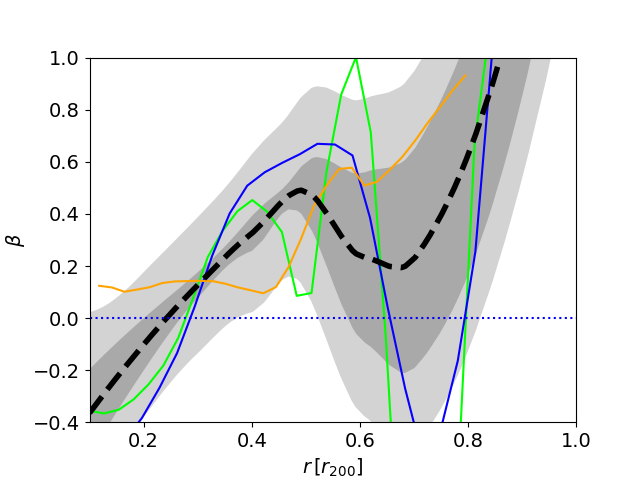}
    \includegraphics[width=0.47\textwidth]{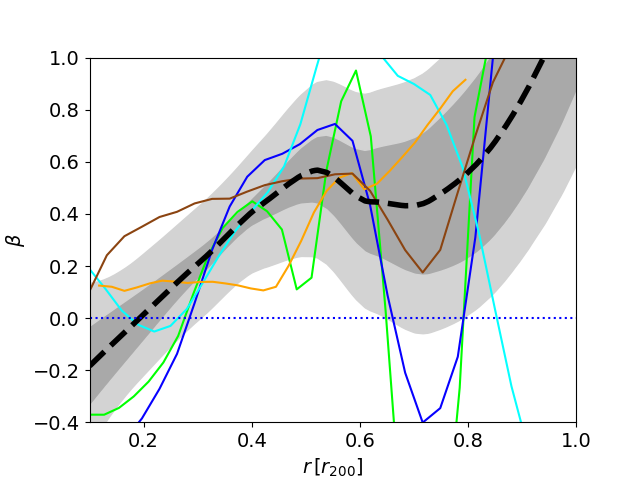}
  \caption{Left: $\beta(r)$ using only the selected 3 sectors which are classified as ``Relaxed'' in \citep{urban2014}. Right: $\beta(r)$ using the 5 equilibrated sectors according to section \ref{sec:perseus} above. In both figures, the black dashed curve shows LOESS fit to the profiles. The dark grey inner band shows the uncertainty of the mean $\frac{\sigma}{\sqrt{N}}$, where $\sigma$ is the spread as obtained through LOESS generalized cross validation, and $N$ is the number of sectors included. The light grey outer band shows the added uncertainty from spread of the RHS of figure \ref{fig:beta}, i.e. the uncertainty from the measuring technique itself. For each sector, multiple $\beta$ profiles are calculated via the analytical framework and MC bootstrap procedure described in the paper. Slight variations upon each MC realisation occurs, and so the profiles may be slightly different between realisations.}
  \label{fig:perseus_beta_all}
\end{figure}

\section{Extracting the DM velocity anisotropy in Perseus}

Having prepared a method for data selection in the previous sections,
and determined 3 sets of sectors for the Perseus cluster to investigate, we
are now ready to extract $\beta$ from it.
We use the deprojected observations of $T_{\mathrm{gas}}$ and 
$\rho_{\mathrm{gas}}$ profiles and their
corresponding error-bars to perform a Monte Carlo sampling as input for the 
analysis for each sector i.e. each arm. First
$\frac{\partial{\rm ln}\rho_{\mathrm{gas}}}{\partial{\rm ln}r}$ and $\frac{\partial
ln T_{\mathrm{gas}}}{\partial{\rm ln}r}$ are found at each radial point by 
computing central differences in the interior and first differences at 
the boundaries. These are then used in the hydrostatic
equilibrium equation to calculate the mass profile. This mass profile
is again subjected to a non-parametric LOESS fit in order to smooth out
bumps and wiggles. After subtracting the gas mass, we can directly
generate the DM density profile. The resulting mass profile of Perseus can be seen 
in figure \ref{fig:perseus_mass} using the data from all 8 arms. For each Monte 
Carlo sampling a DM temperature profile is also resampled
based on the smoothing spline $\kappa$ surface and errors of figure \ref{fig:kappa}, 
and the resampled gas temperature profile.
The $\sigma^2_r(r)$ profiles can now be calculated for each sample, 
and hence the $\beta(r)$ profiles. 

The results are shown in the bottom panel of figure~\ref{fig:perseus_mass} for all 8 
arms of Perseus i.e. the full set.
The coloured curves represents the $\beta(r)$ median Monte Carlo 
profiles obtained from the individual sectors of the Perseus data, 
and the black dashed curve shows another LOESS smoothing to these curves
to estimate an overall $\beta$ for the data included. The inner dark grey band shows
the 1$\sigma$ standard error of the mean as obtained via the standard deviation
of the LOESS fit, and the light grey outer bands the additional 1$\sigma$ standard 
error of the mean from of the standard deviation of the $\beta$ obtained from the 
RAMSES mock data as seen in the left side of figure \ref{fig:beta}. We see that the 
the $\beta$ profile ranges from 0 in the inner parts towards 1 at $r_{200}$ where 
uncertainty grows large on the data and the validity of our assumptions, and thus on 
$\beta$. 

\begin{figure}
  \centering
    \includegraphics[width=0.5\textwidth]{{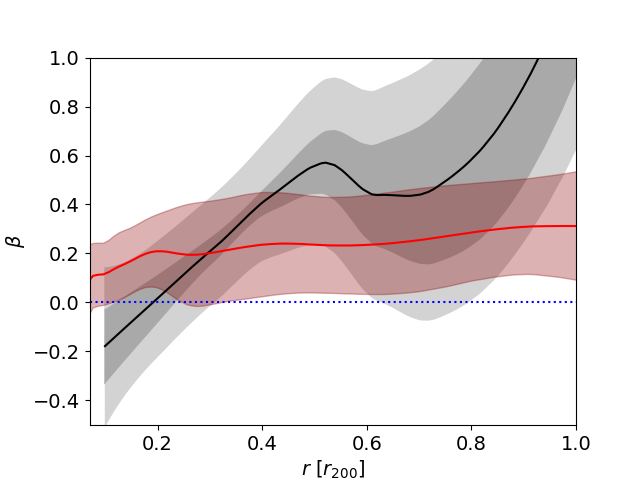}}
  \caption{Comparison of the true $\beta$ profile from RAMSES (red curve) 
  and the estimated $\beta$ profile for Perseus (black curve). The RAMSES profile is comprised 
  of the $\beta$ profiles of the relaxed sectors of the relaxed clusters 
  as also shown in the RHS panel of figure \ref{fig:beta}. The Perseus profile 
  is the same as shown in the bottom panel of figure \ref{fig:perseus_beta_all} 
  for the 5 arms selected through our analysis. Again the dark grey 
  inner band shows the uncertainty of the mean $\frac{\sigma}{\sqrt{N}}$, where $\sigma$ 
  is the standard deviation as obtained through LOESS generalized cross validation, and $N$ 
  is the number of sectors included. The light grey outer band shows the added 
  uncertainty from spread of the RHS of figure \ref{fig:beta}, i.e. the uncertainty 
  from the measuring technique itself.}
  \label{fig:betacompare}
\end{figure}

For the partial sets, we perform the same analysis but include only the 3 sectors
of the Urban et. al. set, and the 5 sectors selected through $T_{\mathrm{W}}$ profiles.
The $\beta$ can be seen in the left and right panel of figure \ref{fig:perseus_beta_all}
respectively. Here, the grey inner bands represent the same as for the full set, but the
outer light grey bands are instead taken from the standard deviation of the right hand side
panel of figure \ref{fig:beta}. We see especially for the set chosen here
that $\beta$ is different from 0 between 0.3$r_{200}$ and 0.6$r_{200}$ beyond its standard
error. Including all the error-bars, we have here found indications
that the velocity anisotropy in Perseus is of the order
\begin{equation}
\beta_{r=0.5r_{200}} = 0.5 \pm 0.1 \pm 0.2 \, , 
\end{equation}
where the error-bars are from: variations within the Perseus cluster
sectors, and the added scatter from the hydrostatic equilibrium
technique itself as applied on each individual arm. This takes the 
optimistic stand that sectors within a galaxy cluster are completely 
uncorrelated measurements of $\beta$. Taking the more
pessimistic viewpoint that sectors within a single cluster are completely 
correlated yields $\beta_{r=0.5r_{200}} = 0.5 \pm 0.1 \pm 0.3$. This includes 
uncertainty from temperature measurements and uncertainty in $\kappa$. 
From around 0.6$\,r_{200}$ and up to 0.8$r_{200}$ the
estimate of $\beta$ is consistent with 0, and beyond that the error grows as the 
assumptions of hydrostatic equilibrium breaks down even for the selected 
clusters and sectors within them. This in fact is already seen in the
measured massprofiles of the RAMSES clusters, where the hydrostatic method
has large errorbars at these large radii. Generally $\beta$ tends to increase
with increasing radius. At $r < 0.2r_{200}$ the results are 
statistically consistent with $\beta = 0$, but if the decrease extends to lower 
radii, it could be interpreted as an effect of the
brightest central galaxy making orbits more tangentially biased. However, this 
effect should not be visible at the scales examined in this work \citep{host2011}.
Figure \ref{fig:betacompare} compares the Perseus estimate
to the $\beta$ profile of the chosen sectors of the RAMSES clusters, and again we see,
for this single Perseus estimate, that observation and simulation 
agrees within $r_{200}$.

\begin{figure}
  \centering
    \includegraphics[width=0.5\textwidth]{{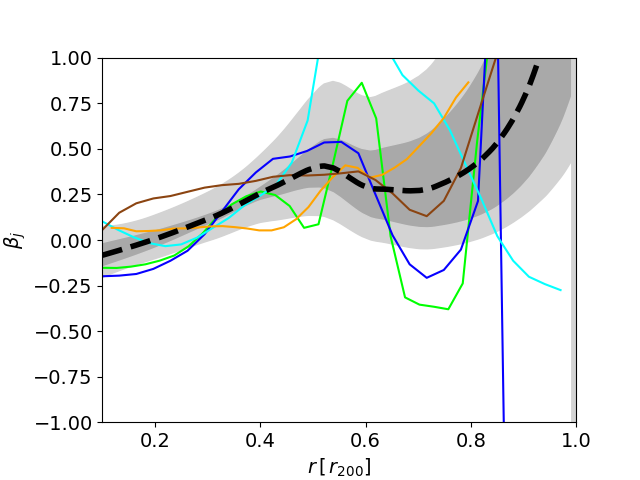}}
  \caption{The non-standard and symmetrized $\beta_j$, which can only assume 
  values between $-1$ and $+1$, plotted for the 5 sectors of Perseus selected in the 
  present work through $T_{\mathrm{W}}$. The black dashed line shows the LOESS fit to the 
  five profiles. The inner and outer grey bands are the 
  standard errors of the mean similarly to the ones of figure \ref{fig:perseus_mass}.}
  \label{fig:betasym}
\end{figure}

It is worth keeping in mind that $\beta$ in principle could take on
any value between $+1$ and $-\infty$. The simulated values of $\beta$
from these 51 RAMSES clusters are about 0.25 at $r=0.5r_{200}$ (with a
1$\sigma$ spread about 0.2 among the 51 clusters), and at $r \sim
r_{200}$ it is about 0.35 (with a dispersion about $^{+0.2}_{-0.4}$).
Upon comparing the measured $\beta$ with the $\beta$ of the simulated
clusters in figure \ref{fig:beta} and figure \ref{fig:perseus_beta_all}
respectively we see that they are in reasonable agreement at least
until 0.7$r_{200}$. Another way to visualize the velocity anisotropy
parameter is through joining $\sigma_r^2$ and $\sigma_t^2$ in the construct
\begin{equation}
\beta_{j} = \frac{\sigma_r^2-\sigma_t^2}{\sigma_r^2+\sigma_t^2}=\frac{\beta}{2-\beta},
\end{equation}
which ranges from -1 to 1. Figure \ref{fig:betasym} shows precisely this
quantity for the 5 sectors of the Perseus cluster selected here, where
the $\beta_j$ profile with its standard error of the mean is seen to be 
non-trivial below 0.6$r_{200}$.

It should be noted, that in spite of removing data from the analysis,
the Perseus resulting $\beta$ is comparable to that of the one with the
full data both in terms of the mean curve and the error. Perseus is
itself a virialized cluster, and thus expectations of bettering the
beta estimate remarkably with this data selection technique are low.
However, as multiple clusters are discovered and analyzed through the same
technique, our results from the RAMSES simulation show that it is 
possible to better the uncertainty in $\beta$ measurements by conducting
data selection of the type outlined here. H\o st et al. obtained 
estimates of $\beta$ within $r_{2500}$ for a stack of 16 clusters
observed in X-ray \citep{host2009}, whereas here we analyze just one single cluster.
In the future we hope to include future cluster X-ray observations to 
high radii to further bring uncertainties down.

\section{Conclusion}

The dark matter velocity anisotropy contains information on 
the dynamics of dark matter in equilibrated structures. By combining the
gas equation from hydrostatic equilibrium and the DM equation i.e. the 
Jeans equation with input from a numerical cosmological simulation that includes
the baryonic component, we are
able to test the consistency of the velocity anisotropy measure for Perseus 
with the dark matter model employed in this simulation. We find that the 
velocity anisotropy of Perseus is consistent with that of the cosmological 
model employing a $\Lambda$CDM cosmology, lending support to the cold and
collisionless nature of DM in galaxy clusters. Our analysis of the Perseus 
data agrees with previous estimates on the velocity anisotropy. Previous studies
employ the strength of a catalogue of 16 galaxy clusters. However, since deprojected
gas profiles are requirement of the analysis the results were within 
0.85$\,r_{2500}$. The quality and radial extent of the Perseus data allows us 
to probe the consistency of the DM model towards the virial radius. By 
analyzing and including only sectors of the cluster that displays a well behaved
radial X-ray signal, we show that we in simulation are able to put better constraints
on the velocity anisotropy estimates, however for Perseus we are only able to get meaningful
constraints towards 0.6$r_{200}$, which is still a large improvement of about a
factor of 3 compared to previous work.

The method comes with some caveats: A relation $\kappa$ from numerical simulations between 
the gas temperature and the dark matter total velocity dispersion i.e. the ``DM temperature''
is used to calibrate the measurement of the velocity anisotropy. The velocity anisotropy measure of a
galaxy cluster in observation is only ever as good as the $\kappa$ that it is
calibrated against. Since we assume a DM model in all cosmological simulations
we at best are able to measure velocity anisotropy as relying on the assumptions
of the simulation. Therefore, our velocity anisotropy measurement should be viewed as 
a check of consistency with the model employed in the estimation of the total DM velocity dispersion.
Furthermore, the Perseus data comprises just a single galaxy cluster. To obtain better constraints on the velocity anisotropy measure is a statistical challenge, and even a couple
of galaxy cluster data sets of the same quality would strengthen the analysis greatly.

Our final remarks concern future analyses. Above we have made the
assumption of hydrostatic equilibrium. It is well known that departure
from hydrostatic equilibrium impacts the mass determinations, see
e.g. \cite{nelson2014} for a list of references, and also
that the velocity anisotropy directly or through mass profiles may depend on orientation
 \citep{sparre2012,wojtak2013,svensmark2015}. We also assume sphericity of the cluster
which have previously been found to impact mass estimates of clusters. Perseus was chosen
because it does have a relaxed appearance, and we have
chosen only the 5 most relaxed arms. Furthermore, at the
moment the large scatter in $\kappa$ leads to large error-bars of $\beta$.  By
analysing cones in numerical simulated clusters, for instance
separating according to differences in temperature profiles such as
cool-core (CC) and non-CC cones, one might be able to reduce scatter
in $\kappa$, and hence obtain smaller error-bars of the DM $\beta$.

Few alternative methods to estimate the DM
velocity anisotropy exits \citep{lemze2012,mamon2013}.
Future analyses which would 
improve on the method discussed here, could be forced to
attempt to improve on the mass determination by including
complementary observation e.g. from lensing or the SZ effect
 \citep{kneib96,stark17}. Here one
will, however, have to deal with the difficult systematic effects when
combining such different observational techniques.

\section*{Acknowledgements}

It is a pleasure thanking Ondrej Urban for providing the data from
Perseus.  We thank Adam Mantz, Aurora Simionescu and Ondrej Urban for
constructive suggestions. SHH wishes to thank Christoffer Bruun-Schmidt, 
Beatriz Soret and Lasse Alb\ae k for discussions. This project is partially 
funded by the Danish council for independent research under the project
``Fundamentals of Dark Matter Structures'', DFF - 6108-00470.

\bibliography{mybib}{}

\begin{thebibliography}{}
\makeatletter
\relax
\def\mn@urlcharsother{\let\do\@makeother \do\$\do\&\do\#\do\^\do\_\do\%\do\~}
\def\mn@doi{\begingroup\mn@urlcharsother \@ifnextchar [ {\mn@doi@}
  {\mn@doi@[]}}
\def\mn@doi@[#1]#2{\def\@tempa{#1}\ifx\@tempa\@empty \href
  {http://dx.doi.org/#2} {doi:#2}\else \href {http://dx.doi.org/#2} {#1}\fi
  \endgroup}
\def\mn@eprint#1#2{\mn@eprint@#1:#2::\@nil}
\def\mn@eprint@arXiv#1{\href {http://arxiv.org/abs/#1} {{\tt arXiv:#1}}}
\def\mn@eprint@dblp#1{\href {http://dblp.uni-trier.de/rec/bibtex/#1.xml}
  {dblp:#1}}
\def\mn@eprint@#1:#2:#3:#4\@nil{\def\@tempa {#1}\def\@tempb {#2}\def\@tempc
  {#3}\ifx \@tempc \@empty \let \@tempc \@tempb \let \@tempb \@tempa \fi \ifx
  \@tempb \@empty \def\@tempb {arXiv}\fi \@ifundefined
  {mn@eprint@\@tempb}{\@tempb:\@tempc}{\expandafter \expandafter \csname
  mn@eprint@\@tempb\endcsname \expandafter{\@tempc}}}

\bibitem[\protect\citeauthoryear{{ATLAS Collaboration}, Aad, Abbott, Abdallah
  \& Abdel~Khalek}{{ATLAS Collaboration} et~al.}{2015}]{aad2015}
{ATLAS Collaboration} Aad G.,  Abbott B.,  Abdallah J.,   Abdel~Khalek S.,
  2015, \mn@doi [The European Physical Journal C]
  {10.1140/epjc/s10052-015-3306-z}, 75, 92

\bibitem[\protect\citeauthoryear{Agnes et~al.,}{Agnes et~al.}{2015}]{agnes2015}
Agnes P.,  et~al., 2015, \mn@doi [Physics Letters B]
  {https://doi.org/10.1016/j.physletb.2015.03.012}, 743, 456

\bibitem[\protect\citeauthoryear{Albæk, Hansen, Martizzi, Moore  \&
  Teyssier}{Albæk et~al.}{2017}]{albaek2017}
Albæk L.,  Hansen S.~H.,  Martizzi D.,  Moore B.,   Teyssier R.,  2017,
  \mn@doi [Monthly Notices of the Royal Astronomical Society]
  {10.1093/mnras/stx2139}, 472, 3486

\bibitem[\protect\citeauthoryear{Amorisco, Zavala  \& de Boer}{Amorisco
  et~al.}{2014}]{amorisco2014}
Amorisco N.~C.,  Zavala J.,   de Boer T. J.~L.,  2014, The Astrophysical
  Journal Letters, 782, L39

\bibitem[\protect\citeauthoryear{Armitage, Kay, Barnes, Bahé  \&
  Dalla Vecchia}{Armitage et~al.}{2018}]{armitage2019}
Armitage T.~J.,  Kay S.~T.,  Barnes D.~J.,  Bahé Y.~M.,   Dalla Vecchia C.,
  2018, \mn@doi [Monthly Notices of the Royal Astronomical Society]
  {10.1093/mnras/sty2921}, 482, 3308

\bibitem[\protect\citeauthoryear{{Benitez-Llambay}, {Frenk}, {Ludlow}  \&
  {Navarro}}{{Benitez-Llambay} et~al.}{2018}]{benitez-llambay2018}
{Benitez-Llambay} A.,  {Frenk} C.~S.,  {Ludlow} A.~D.,   {Navarro} J.~F.,
  2018, arXiv e-prints, \href
  {https://ui.adsabs.harvard.edu/\#abs/2018arXiv181004186B} {p.
  arXiv:1810.04186}

\bibitem[\protect\citeauthoryear{{Binney} \& {Tremaine}}{{Binney} \&
  {Tremaine}}{2008}]{bt87}
{Binney} J.,  {Tremaine} S.,  2008, {Galactic Dynamics: Second Edition}.
Princeton University Press

\bibitem[\protect\citeauthoryear{B\oe{}hm, Riazuelo, Hansen  \&
  Schaeffer}{B\oe{}hm et~al.}{2002}]{bloehm2002}
B\oe{}hm C.,  Riazuelo A.,  Hansen S.~H.,   Schaeffer R.,  2002, \mn@doi [Phys.
  Rev. D] {10.1103/PhysRevD.66.083505}, 66, 083505

\bibitem[\protect\citeauthoryear{{Bose} et~al.,}{{Bose}
  et~al.}{2018}]{bose2018}
{Bose} S.,  et~al., 2018, arXiv e-prints, \href
  {https://ui.adsabs.harvard.edu/\#abs/2018arXiv181003635B} {p.
  arXiv:1810.03635}

\bibitem[\protect\citeauthoryear{Brinckmann, Zavala, Rapetti, Hansen  \&
  Vogelsberger}{Brinckmann et~al.}{2018}]{brinckmann2018}
Brinckmann T.,  Zavala J.,  Rapetti D.,  Hansen S.~H.,   Vogelsberger M.,
  2018, \mn@doi [Monthly Notices of the Royal Astronomical Society]
  {10.1093/mnras/stx2782}, 474, 746

\bibitem[\protect\citeauthoryear{Bullock, Oñorbe, Boylan-Kolchin, Hopkins,
  Kereš, Faucher-Giguère, Quataert  \& Murray}{Bullock
  et~al.}{2015}]{bullock2015}
Bullock J.~S.,  Oñorbe J.,  Boylan-Kolchin M.,  Hopkins P.~F.,  Kereš D.,
  Faucher-Giguère C.-A.,  Quataert E.,   Murray N.,  2015, \mn@doi [Monthly
  Notices of the Royal Astronomical Society] {10.1093/mnras/stv2072}, 454, 2092

\bibitem[\protect\citeauthoryear{Clowe, Bradač, Gonzalez, Markevitch, Randall,
  Jones  \& Zaritsky}{Clowe et~al.}{2006}]{clowe2006}
Clowe D.,  Bradač M.,  Gonzalez A.~H.,  Markevitch M.,  Randall S.~W.,  Jones
  C.,   Zaritsky D.,  2006, The Astrophysical Journal Letters, 648, L109

\bibitem[\protect\citeauthoryear{Di~Cintio, Tremmel, Governato, Pontzen,
  Zavala, Bastidas~Fry, Brooks  \& Vogelsberger}{Di~Cintio
  et~al.}{2017}]{dicintio2017}
Di~Cintio A.,  Tremmel M.,  Governato F.,  Pontzen A.,  Zavala J.,
  Bastidas~Fry A.,  Brooks A.,   Vogelsberger M.,  2017, \mn@doi [Monthly
  Notices of the Royal Astronomical Society] {10.1093/mnras/stx1043}, 469, 2845

\bibitem[\protect\citeauthoryear{{Dutton}, {Macci{\`o}}, {Buck}, {Dixon},
  {Blank}  \& {Obreja}}{{Dutton} et~al.}{2018}]{dutton2018}
{Dutton} A.~A.,  {Macci{\`o}} A.~V.,  {Buck} T.,  {Dixon} K.~L.,  {Blank} M.,
  {Obreja} A.,  2018, arXiv e-prints, \href
  {https://ui.adsabs.harvard.edu/\#abs/2018arXiv181110625D} {p.
  arXiv:1811.10625}

\bibitem[\protect\citeauthoryear{{Faham}}{{Faham}}{2014}]{faham2014}
{Faham} C.,  2014, preprint, \href
  {http://adsabs.harvard.edu/abs/2014arXiv1405.5906F} {} (\mn@eprint {arXiv}
  {1405.5906})

\bibitem[\protect\citeauthoryear{Falco, Mamon, Wojtak, Hansen  \&
  Gottlöber}{Falco et~al.}{2013}]{falco2013}
Falco M.,  Mamon G.~A.,  Wojtak R.,  Hansen S.~H.,   Gottlöber S.,  2013,
  \mn@doi [Monthly Notices of the Royal Astronomical Society]
  {10.1093/mnras/stt1768}, 436, 2639

\bibitem[\protect\citeauthoryear{{Fitts} et~al.,}{{Fitts}
  et~al.}{2018}]{fitts2018}
{Fitts} A.,  et~al., 2018, arXiv e-prints, \href
  {https://ui.adsabs.harvard.edu/\#abs/2018arXiv181111791F} {p.
  arXiv:1811.11791}

\bibitem[\protect\citeauthoryear{Gifford \& Miller}{Gifford \&
  Miller}{2013}]{gifford2013}
Gifford D.,  Miller C.~J.,  2013, \mn@doi [The Astrophysical Journal]
  {10.1088/2041-8205/768/2/l32}, 768, L32

\bibitem[\protect\citeauthoryear{Gilmore, Wilkinson, Wyse, Kleyna, Koch, Evans
  \& Grebel}{Gilmore et~al.}{2007}]{gilmore2007}
Gilmore G.,  Wilkinson M.~I.,  Wyse R. F.~G.,  Kleyna J.~T.,  Koch A.,  Evans
  N.~W.,   Grebel E.~K.,  2007, The Astrophysical Journal, 663, 948

\bibitem[\protect\citeauthoryear{González-Samaniego
  et~al.,}{González-Samaniego et~al.}{2017}]{gonzales-sameniego2017}
González-Samaniego A.,  et~al., 2017, \mn@doi [Monthly Notices of the Royal
  Astronomical Society] {10.1093/mnras/stx2253}, 472, 2945

\bibitem[\protect\citeauthoryear{Hansen}{Hansen}{2009}]{hansen2009}
Hansen S.~H.,  2009, The Astrophysical Journal, 694, 1250

\bibitem[\protect\citeauthoryear{{Hansen, S. H.} \& {Piffaretti, R.}}{{Hansen,
  S. H.} \& {Piffaretti, R.}}{2007}]{hansen2007}
{Hansen, S. H.} {Piffaretti, R.} 2007, \mn@doi [A\&A]
  {10.1051/0004-6361:20078656}, 476, L37

\bibitem[\protect\citeauthoryear{Hansen, Macció, Romano-Diaz, Hoffman,
  Brüggen, Scannapieco  \& Stinson}{Hansen et~al.}{2011}]{hansen2011}
Hansen S.~H.,  Macció A.~V.,  Romano-Diaz E.,  Hoffman Y.,  Brüggen M.,
  Scannapieco E.,   Stinson G.~S.,  2011, The Astrophysical Journal, 734, 62

\bibitem[\protect\citeauthoryear{Hinshaw et~al.,}{Hinshaw
  et~al.}{2013}]{hinshaw2013}
Hinshaw G.,  et~al., 2013, The Astrophysical Journal Supplement Series, 208, 19

\bibitem[\protect\citeauthoryear{Host \& Hansen}{Host \&
  Hansen}{2011}]{host2011}
Host O.,  Hansen S.~H.,  2011, \mn@doi [The Astrophysical Journal]
  {10.1088/0004-637x/736/1/52}, 736, 52

\bibitem[\protect\citeauthoryear{Host, Hansen, Piffaretti, Morandi, Ettori, Kay
   \& Valdarnini}{Host et~al.}{2009}]{host2009}
Host O.,  Hansen S.~H.,  Piffaretti R.,  Morandi A.,  Ettori S.,  Kay S.~T.,
  Valdarnini R.,  2009, The Astrophysical Journal, 690, 358

\bibitem[\protect\citeauthoryear{Kay, Da~Silva, Aghanim, Blanchard, Liddle,
  Puget, Sadat  \& Thomas}{Kay et~al.}{2007}]{kay2007}
Kay S.~T.,  Da~Silva A.~C.,  Aghanim N.,  Blanchard A.,  Liddle A.~R.,  Puget
  J.-L.,  Sadat R.,   Thomas P.~A.,  2007, \mn@doi [Monthly Notices of the
  Royal Astronomical Society] {10.1111/j.1365-2966.2007.11605.x}, 377, 317

\bibitem[\protect\citeauthoryear{Kneib \& Natarajan}{Kneib \&
  Natarajan}{1996}]{kneib96}
Kneib J.-P.,  Natarajan P.,  1996, \mn@doi [Monthly Notices of the Royal
  Astronomical Society] {10.1093/mnras/283.3.1031}, 283, 1031

\bibitem[\protect\citeauthoryear{Lemze et~al.,}{Lemze et~al.}{2012}]{lemze2012}
Lemze D.,  et~al., 2012, \mn@doi [The Astrophysical Journal]
  {10.1088/0004-637x/752/2/141}, 752, 141

\bibitem[\protect\citeauthoryear{Liu, Slatyer  \& Zavala}{Liu
  et~al.}{2016}]{liu2016}
Liu H.,  Slatyer T.~R.,   Zavala J.,  2016, \mn@doi [Phys. Rev. D]
  {10.1103/PhysRevD.94.063507}, 94, 063507

\bibitem[\protect\citeauthoryear{{Lowette} \& {for the CMS
  Collaboration}}{{Lowette} \& {for the CMS Collaboration}}{2014}]{lowette2014}
{Lowette} S.,  {for the CMS Collaboration} 2014, preprint, \href
  {http://adsabs.harvard.edu/abs/2014arXiv1410.3762L} {} (\mn@eprint {arXiv}
  {1410.3762})

\bibitem[\protect\citeauthoryear{Mamon, Boué  \& Biviano}{Mamon
  et~al.}{2013}]{mamon2013}
Mamon G.~A.,  Boué G.,   Biviano A.,  2013, \mn@doi [Monthly Notices of the
  Royal Astronomical Society] {10.1093/mnras/sts565}, 429, 3079

\bibitem[\protect\citeauthoryear{Markevitch \& Vikhlinin}{Markevitch \&
  Vikhlinin}{2007}]{markevitch2007}
Markevitch M.,  Vikhlinin A.,  2007, \mn@doi [Physics Reports]
  {https://doi.org/10.1016/j.physrep.2007.01.001}, 443, 1

\bibitem[\protect\citeauthoryear{{Martizzi} \& {Agrusa}}{{Martizzi} \&
  {Agrusa}}{2016}]{martizzi2016}
{Martizzi} D.,  {Agrusa} H.,  2016, arXiv e-prints, \href
  {https://ui.adsabs.harvard.edu/\#abs/2016arXiv160804388M} {p.
  arXiv:1608.04388}

\bibitem[\protect\citeauthoryear{Martizzi, Mohammed, Teyssier  \&
  Moore}{Martizzi et~al.}{2014}]{martizzi2014}
Martizzi D.,  Mohammed I.,  Teyssier R.,   Moore B.,  2014, \mn@doi [Monthly
  Notices of the Royal Astronomical Society] {10.1093/mnras/stu440}, 440, 2290

\bibitem[\protect\citeauthoryear{Navarro et~al.,}{Navarro
  et~al.}{2010}]{navarro2010}
Navarro J.~F.,  et~al., 2010, \mn@doi [Monthly Notices of the Royal
  Astronomical Society] {10.1111/j.1365-2966.2009.15878.x}, 402, 21

\bibitem[\protect\citeauthoryear{Nelson, Lau  \& Nagai}{Nelson
  et~al.}{2014}]{nelson2014}
Nelson K.,  Lau E.~T.,   Nagai D.,  2014, \mn@doi [The Astrophysical Journal]
  {10.1088/0004-637x/792/1/25}, 792, 25

\bibitem[\protect\citeauthoryear{{Planck Collaboration} et~al.,}{{Planck
  Collaboration} et~al.}{2018}]{planck18}
{Planck Collaboration} et~al., 2018, arXiv e-prints, \href
  {https://ui.adsabs.harvard.edu/\#abs/2018arXiv180706209P} {p.
  arXiv:1807.06209}

\bibitem[\protect\citeauthoryear{{Pointecouteau, E.}, {Arnaud, M.}  \& {Pratt,
  G. W.}}{{Pointecouteau, E.} et~al.}{2005}]{pointecouteau2005}
{Pointecouteau, E.} {Arnaud, M.}  {Pratt, G. W.} 2005, \mn@doi [A\&A]
  {10.1051/0004-6361:20042569}, 435, 1

\bibitem[\protect\citeauthoryear{Salucci, Lapi, Tonini, Gentile, Yegorova  \&
  Klein}{Salucci et~al.}{2007}]{salucci2007}
Salucci P.,  Lapi A.,  Tonini C.,  Gentile G.,  Yegorova I.,   Klein U.,  2007,
  \mn@doi [Monthly Notices of the Royal Astronomical Society]
  {10.1111/j.1365-2966.2007.11696.x}, 378, 41

\bibitem[\protect\citeauthoryear{Santos-Santos, Di~Cintio, Brook, Macciò,
  Dutton  \& Domínguez-Tenreiro}{Santos-Santos
  et~al.}{2017}]{santos-santos2017}
Santos-Santos I.~M.,  Di~Cintio A.,  Brook C.~B.,  Macciò A.,  Dutton A.,
  Domínguez-Tenreiro R.,  2017, \mn@doi [Monthly Notices of the Royal
  Astronomical Society] {10.1093/mnras/stx2660}, 473, 4392

\bibitem[\protect\citeauthoryear{Sarazin}{Sarazin}{1986}]{sarazin1986}
Sarazin C.~L.,  1986, \mn@doi [Rev. Mod. Phys.] {10.1103/RevModPhys.58.1}, 58,
  1

\bibitem[\protect\citeauthoryear{Scrucca}{Scrucca}{2011}]{msir2011}
Scrucca L.,  2011, Computational Statistics \& Data Analysis, 5, 3010

\bibitem[\protect\citeauthoryear{Simionescu et~al.,}{Simionescu
  et~al.}{2011}]{simionescu2011}
Simionescu A.,  et~al., 2011, \mn@doi [Science] {10.1126/science.1200331}, 331,
  1576

\bibitem[\protect\citeauthoryear{Simionescu et~al.,}{Simionescu
  et~al.}{2012}]{simionescu2012}
Simionescu A.,  et~al., 2012, \mn@doi [The Astrophysical Journal]
  {10.1088/0004-637x/757/2/182}, 757, 182

\bibitem[\protect\citeauthoryear{Sparre \& Hansen}{Sparre \&
  Hansen}{2012}]{sparre2012}
Sparre M.,  Hansen S.~H.,  2012, Journal of Cosmology and Astroparticle
  Physics, 2012, 042

\bibitem[\protect\citeauthoryear{{Stark}, {Miller}  \& {Halenka}}{{Stark}
  et~al.}{2017}]{stark17}
{Stark} A.,  {Miller} C.~J.,   {Halenka} V.,  2017, arXiv e-prints, \href
  {https://ui.adsabs.harvard.edu/\#abs/2017arXiv171110018S} {p.
  arXiv:1711.10018}

\bibitem[\protect\citeauthoryear{Svensmark, Hansen  \& Wojtak}{Svensmark
  et~al.}{2015}]{svensmark2015}
Svensmark J.,  Hansen S.~H.,   Wojtak R.,  2015, \mn@doi [Monthly Notices of
  the Royal Astronomical Society] {10.1093/mnras/stu2686}, 448, 1644

\bibitem[\protect\citeauthoryear{{Teyssier, R.}}{{Teyssier,
  R.}}{2002}]{teyssier2002}
{Teyssier, R.} 2002, \mn@doi [A\&A] {10.1051/0004-6361:20011817}, 385, 337

\bibitem[\protect\citeauthoryear{Teyssier, Pontzen, Dubois  \& Read}{Teyssier
  et~al.}{2013}]{teyssier2013}
Teyssier R.,  Pontzen A.,  Dubois Y.,   Read J.~I.,  2013, \mn@doi [Monthly
  Notices of the Royal Astronomical Society] {10.1093/mnras/sts563}, 429, 3068

\bibitem[\protect\citeauthoryear{Urban et~al.,}{Urban et~al.}{2014}]{urban2014}
Urban O.,  et~al., 2014, \mn@doi [Monthly Notices of the Royal Astronomical
  Society] {10.1093/mnras/stt2209}, 437, 3939

\bibitem[\protect\citeauthoryear{Vikhlinin, Kravtsov, Forman, Jones,
  Markevitch, Murray  \& Speybroeck}{Vikhlinin et~al.}{2006}]{vikhlinin2006}
Vikhlinin A.,  Kravtsov A.,  Forman W.,  Jones C.,  Markevitch M.,  Murray
  S.~S.,   Speybroeck L.~V.,  2006, The Astrophysical Journal, 640, 691

\bibitem[\protect\citeauthoryear{Wang}{Wang}{2010}]{fANCOVA2010}
Wang X.-F.,  2010, fANCOVA: Nonparametric Analysis of Covariance.
\url {https://CRAN.R-project.org/package=fANCOVA}

\bibitem[\protect\citeauthoryear{Wetzel, Hopkins, hoon Kim,
  Faucher-Gigu{\`{e}}re, Kere{\v{s}}  \& Quataert}{Wetzel
  et~al.}{2016}]{wetzel2016}
Wetzel A.~R.,  Hopkins P.~F.,  hoon Kim J.,  Faucher-Gigu{\`{e}}re C.-A.,
  Kere{\v{s}} D.,   Quataert E.,  2016, \mn@doi [The Astrophysical Journal]
  {10.3847/2041-8205/827/2/l23}, 827, L23

\bibitem[\protect\citeauthoryear{{Wheeler} et~al.,}{{Wheeler}
  et~al.}{2018}]{wheeler2018}
{Wheeler} C.,  et~al., 2018, arXiv e-prints, \href
  {https://ui.adsabs.harvard.edu/\#abs/2018arXiv181202749W} {p.
  arXiv:1812.02749}

\bibitem[\protect\citeauthoryear{Wojtak, Gottlöber  \& Klypin}{Wojtak
  et~al.}{2013}]{wojtak2013}
Wojtak R.,  Gottlöber S.,   Klypin A.,  2013, \mn@doi [Monthly Notices of the
  Royal Astronomical Society] {10.1093/mnras/stt1113}, 434, 1576

\bibitem[\protect\citeauthoryear{Zavala, Vogelsberger  \& Walker}{Zavala
  et~al.}{2013}]{zavala2013}
Zavala J.,  Vogelsberger M.,   Walker M.~G.,  2013, \mn@doi [Monthly Notices of
  the Royal Astronomical Society: Letters] {10.1093/mnrasl/sls053}, 431, L20

\makeatother
\end{thebibliography}
\bibliographystyle{mnras}
\bsp	
\label{lastpage}
\end{document}